\documentclass[seceq]{ptptex}

\usepackage{amsmath,amsfonts,latexsym,amssymb}
\usepackage{verbatim,amsthm,graphicx}
\usepackage{wrapft}
\usepackage{mathrsfs, amsmath}

\def\refsec#1{\S \ref{sec: #1}}

\def\refeq#1{(\ref{eq: #1})}

\def\Eq#1{\begin{equation} #1 \end{equation}}
\def\Eqr#1{\begin{eqnarray} #1 \end{eqnarray}}
\def\dfrac#1#2{\displaystyle\frac{#1}{#2}}
\def\lineq{\hspace{0.35em}\raisebox{0.4ex}{$<$}\hspace{-0.75em}\raisebox{-.7ex}{$\sim$}\hspace{0.3em}}

\def\p#1{\partial_{#1}}
\def\Ord#1{{\rm O}(#1)}
\def\Ft{\p t F_0}
\def\Fr{\p r F_0}
\def\Fo{\p\omega F_0}
\def\Fm{\p\mu F_0}
\def\Fy{\omega\Fo}
\def\Fyy{(\omega\p\omega)^2 F_0}
\def\Fyr{\omega\p\omega\Fr}
\def\Fym{\omega\p\omega\Fm}
\def\Frr{\partial^2_r F_0}
\def\Frm{\p r \Fm}
\def\Fmm{\partial^2_{\mu}F_0}
\def\ape{a_{\perp}}
\def\apa{a_{\scriptscriptstyle /\! /}}
\def\Hpe{H_{\perp}}
\def\Hpa{H_{\scriptscriptstyle \! /\! /}}

\def\eU#1#2{e^{-U(#1, #2)}}
\def\e2U#1#2{e^{-2U(#1, #2)}}

\def\t{\tilde }
\def\bm#1{\boldsymbol{#1}}

\begin{document}

\title{Analytic Formulae for the Off-Center CMB Anisotropy \\
		in a General Spherically Symmetric Universe}

\author{Hideo Kodama\footnote{Email address: hideo.kodama@kek.jp}$^{(a)}$,
Keiki Saito\footnote{Email address: saitok@post.kek.jp}$^{(b)}$
and Akihiro Ishibashi\footnote{Email address: akihiro.ishibashi@kek.jp}$^{(a)}$}

\inst{$^{(a)}$KEK Theory Center, Institute of Particle and Nuclear Studies, KEK, \\
1-1 Oho, Tsukuba, Ibaraki, 305-0801, Japan\\
$^{(b)}$Department of Particles and Nuclear Physics, \\
The Graduate University for Advanced Studies (SOKENDAI), \\
1-1 Oho, Tsukuba, Ibaraki 305-0801, Japan}

\abst{
The local void model has recently attracted considerable attention because 
it can explain the apparent accelerated expansion of the present
universe without introducing dark energy. However, in order to
justify this model as an alternative 
to the standard $\Lambda$CDM cosmology, 
the model should be tested by various observations, such as the CMB
temperature anisotropy, besides the distance-redshift relation of
SNIa. For this purpose, we derive analytic formulae 
for the dipole and quadrupole moments of the CMB temperature anisotropy  
that hold for any spherically symmetric universe model and 
can be used to compare consequences of such a model
with observations of the CMB temperature anisotropy rigorously.  
We check that our formulae are consistent with the numerical studies 
previously made for the CMB temperature anisotropy in the void model. 
We also update the constraints concerning the location of 
the observers in the void model by applying our analytic dipole formula 
with the latest WMAP data.  
}

\maketitle

\section{Introduction}

In modern cosmology, it is commonly assumed that our universe 
be isotropic and homogeneous on 
large scales and accordingly be described by the 
Friedmann-Lema$\hat{\mbox{\i}}$tre-Robertson-Walker (FLRW) metric 
as a first approximation. 
Then, together with consequences of various cosmological observations
such as the power spectrum of Cosmic Microwave Background (CMB) temperature anisotropy, 
the distance-redshift relation of type Ia supernovae (SNIa)
indicates that the expansion of the present universe is accelerated. 
We are then led to introduce 
``dark energy,'' which has negative pressure and behaves just like
a positive cosmological constant, thus having the standard $\Lambda$CDM 
model. 
However, there does not appear to be any satisfactory theory that can 
naturally explain the origin of dark energy and its magnitude
required by observations. It is therefore tempting to seek alternative
explanations of the apparent cosmic acceleration, more specifically,
the SNIa luminosity distance-redshift relation, without invoking dark
energy. (For a number of approaches to the dark energy problem, 
see e.g., \citen{CST06} and references therein).

One of such attempts is a ``local void model'' proposed by
Tomita~\cite{Tomita1st,Tomita1st2}, also independently by Celerier~\cite{Celerier}, 
and by Goodwin et al \cite{Goodwin}.
In this model, our universe is no longer assumed to be homogeneous,
having instead an under-dense local void in the surrounding overdense 
universe.\footnote{ 
See, e.g., \citen{LTBreview} for earlier work along a similar 
line. For other proposals that also attempt to exploit the effects 
of inhomogeneities to solve the dark energy problem, 
interests and criticisms thereof, see e.g., \citen{Buchert00,Buchert03,Ellis,Rasanen04,Kolb05,Kolb06,Kasai92,Nambu,Kai,Tanimoto,Wiltshire05,Flanagan05,HS05,IW06,KAF06}. 
} 
The isotropic nature of cosmological observations on large scales
is realized by assuming the spherical symmetry and demanding that
we live close to the center of the void. Furthermore, the model is
supposed to contain only ordinary dust like cosmic matter, describing,
say, the CDM component.  
Since such a spacetime can be described by 
the Lema$\hat{\mbox{\i}}$tre-Tolman-Bondi (LTB) metric \cite{L,T,B}, 
we also call this model the ``LTB cosmological model.''\footnote{ 
In fact, the LTB metric has been considered in the cosmological context  
even much before the present form of the dark energy problem was raised 
by the results of SNIa observations. 
For example, an early proposal of the void model constructed by the LTB 
metric was made by Moffat and Tatarski~\cite{MT92,MT95}. 
} 
Since the rate of expansion in the void region is larger than that
in the outer overdense region, this model can account for the observed 
dimming of SNIa luminosity. 
In fact, recent numerical analyses \cite{Tomita2nd,Iguchi,Kasai,AAG,Moffat,Mansouri,Garfinkle,Vanderveld,Biswas,Yoo,GBH,Clifton,Zibin,Alexander} 
have shown that the LTB model can accurately reproduce the SNIa 
distance-redshift relation. For this reason, despite the relinquishment of
the widely accepted Copernican/cosmological principle, the local void/LTB
model has recently attracted considerable
attention.

However, in order to justify the LTB model as a viable alternative to
the standard $\Lambda$CDM model, 
one has to test this model by various observations 
other than the SNIa distance-redshift relation.  
A number of papers for this purpose have appeared recently, studying 
constraints from observations, such as the CMB temperature anisotropy
\cite{AAG,Moffat,TomitaCMB,Alexander}, 
the baryon acoustic oscillation~\cite{Gaztanaga,Zibin,Alexander}, 
the kinematic Sunyaev-Zeldovich effect~\cite{Benson,GBHkSZ}, the galaxy count-redshift relation~\cite{R07,R10,R09_1,R09_2}, etc.
Many LTB models with a small, a few hundreds Mpc size void are already 
ruled out, but at present, the models with a huge, Gpc size void 
still remain to be tested. More details of the current status of 
the LTB models are discussed in \citen{LTBreview}. 
Most of these analyses have been performed for various types of 
LTB models by using numerical methods, and it does not seem to be 
straightforward to compare analyses for each different model 
so as to have a coherent understanding of the results. 
In order to have general consequences of the LTB cosmology and 
systematically examine its viability, it is desirable to develop 
some general, analytic methods that 
can apply, independently of the details of each specific model.


Apart from the quest of alternatives to dark energy, the LTB metric
may also serve as a toy model for getting some insights into
the dynamics and possible observational effects of non-linear
perturbations in the standard $\Lambda$CDM cosmology.
The LTB metric can incorporate a cosmological constant in a straightforward
manner, hence being able to describe, as an exact solution, 
highly non-linear inhomogeneities---almost arbitrary in magnitude,
as long as being spherically symmetric---in the FLRW universe with
dark energy.   
In view of this, it is also worth attempting to derive some analytic
formulae that can be used to make theoretical predictions of the
$\Lambda$-LTB spacetimes and rigorous comparison with cosmological
observations.

The purpose of this paper is to derive analytic formulae for the dipole and quadrupole moments of 
the CMB temperature anisotropy in general 
spherically symmetric inhomogeneous spacetimes, including
the $\Lambda$-LTB spacetime as a particular case. 
In the standard cosmology, basic properties of the CMB temperature 
anisotropy are derived by inspecting perturbations of the Einstein
and Boltzmann equations in the FLRW background universe.
Ideally, it is desirable to do the same thing in the LTB background. 
However, perturbations in a LTB spacetime, let alone those in a general
spherically symmetric spacetime, have not been very well studied mainly
because the perturbation equations are much more involved to solve 
in the spherical but inhomogeneous background, 
though the linear perturbation formulae themselves have long been 
available~\cite{GS79a,GS79b}\footnote{ 
See also for some recent paper relevant to the LTB cosmology~\cite{CTF09}.  
}. 
In this paper, we are not going to deal with perturbations of the LTB 
metric. Instead, we will exploit the key requirement of the LTB cosmology 
that we, observers, are restricted to be around very near the 
center of the spherical symmetry: Namely, we first note that 
the small distance between the symmetry center and an off-center 
observer gives rise to a corresponding deviation in the photon 
distribution function. Then, by taking `Taylor-expansions' of the photon 
distribution function at the center with respect to the deviation, 
we can read off the CMB temperature anisotropy caused by the deviation 
in the photon distribution function. 
By doing so, we can, in principle, construct the $l$-th order multiple 
moment of the CMB temperature anisotropy from the (up to) $l$-th order 
expansion coefficients, with the help of the background null geodesic 
equations and the Boltzmann equation. 
We will do so for the first and second-order expansions to find 
the CMB dipole and quadrupole moments. We also provide the concrete 
expression of the corresponding formulae for the LTB cosmological model. 
Our formulae are then checked to be consistent with the numerical analyses 
of the CMB temperature anisotropy in the LTB model, previously made by
Alnes and Amarzguioui \cite{AA}. 
We then apply our formulae to place the constraint on the distance
between an observer and the symmetry center of the void, by using
the latest Wilkinson Microwave Anisotropy Probe (WMAP) data, 
thereby updating the results of the previous analyses.


In the next section, 
we derive analytic formulae for the CMB temperature anisotropy in the 
most general spherically symmetric spacetime. In \refsec{LTB}, we obtain
analytic formulae for the CMB temperature anisotropy in the LTB model,
and some constraints concerning the position of the observer.
\refsec{summary} is devoted to summary. Appendix discusses 
the regularity of the photon distribution function 
and the behavior of some geometric quantities at the symmetry center.

\section{The CMB temperature anisotropy in  spherically symmetric
 spacetimes} 
\label{sec: sphe. sym.}

In this section, we will derive analytic formulae for the dipole and 
quadrupole moments of the CMB temperature anisotropy. 
We first briefly discuss the null geodesic equations, 
the photon distribution function and its relation 
to the CMB photon temperature in the most general spherically 
symmetric spacetime. 
Next, inspecting the first-order derivatives of 
the photon distribution function at the center of the spherical 
symmetry and using the null geodesic equations, we derive 
the dipole formula for the CMB. Then, taking further 
derivatives, we derive the quadrupole formula of the 
CMB temperature anisotropy in general spherically symmetric spacetimes.

\subsection{The photon distribution function in a general spherically 
symmetric spacetime}

The most general spherically symmetric metric can be written 
in the following form 
\Eq{
ds^2 = -N^2 (t, r) dt^2 + S^2 (t, r) dr^2 
     + R^2 (t, r)(d\theta^2 + \sin^2 \theta d\phi^2). 
\label{eq: spherically sym.} 
}
In this background, the relevant geodesic equations are given by 
\Eqr{
\frac{dp^t}{d\lambda} 
         &=& -\frac{\dot N}{N}(p^t)^2 
             - \frac{2N'}{N}p^t p^r - \frac{S\dot S}{N^2}(p^r)^2 
             - \frac{\dot R}{N^2 R}p_\perp^2, \label{eq: pt}\\
\frac{dp^r}{d\lambda} 
         &=& -\frac{NN'}{S^2}(p^t)^2 - \frac{2\dot S}{S}p^t p^r 
             - \frac{S'}{S}(p^r)^2 + \frac{R'}{S^2 R}p_\perp^2, \label{eq: pr}
}
where $p^\mu = dx^\mu/d\lambda$, $p_\mu p^\mu=0$,  
with $\lambda$ being an affine parameter, 
and $p_\perp$ is given by 
$p_\perp^2 \equiv R^2 \{(p^\theta)^2 + (p^\phi)^2 \sin^2 \theta\}$.   
Here and in the rest of the paper, {\em prime} and {\em dot} denote 
the derivatives with respect to $r$ and $t$, respectively.

We are concerned with the distribution function, $F(x, p)$, for the CMB 
photons that leave the ``last scattering surface'' appropriately defined, 
say, $t=t_i$ hypersurface, in the universe modelled by the above metric 
and that eventually reach an ``observer'' very near the symmetry center. 
We assume that the photon distribution function $F(x, p)$ is 
spherically symmetric, respecting the symmetry of the background 
geometry, so that 
\Eqr{
 F(x, p) = F_0(t,r,\omega,\mu) , 
\label{eq:F:spherical}
} 
where $\omega \equiv p^t$ and $\mu \equiv Sp^r/(N\omega)$. 
Note that from the above geodesic equations, we have 
\Eqr{
\dot{r} &=& \mu\dfrac{N}{S}, \label{eq: dotr}\\
\dot{\omega} &=& -\omega
                  \left(\dfrac{\dot N}{N} 
                       + 2\mu\dfrac{N'}{S} + \mu^2 \dfrac{\dot S}{S} 
                       + (1 - \mu^2 ) \dfrac{\dot R}{R}
                  \right), 
             \label{eq: dotomega}\\
\dot{\mu} &=& (1 - \mu^2 )
              \left\{
                    \dfrac{N}{S}
                    \left(\dfrac{R'}{R} - \frac{N'}{N}\right) 
                   + \mu\left( 
                              \dfrac{\dot R}{R} - \dfrac{\dot S}{S}
                        \right)
              \right\}.   
\label{eq: dotmu} 
}
In particular, it follows from the above equations 
\Eqr{
    \p\omega\dot r = \p\omega\dot{\mu} = 0 ,\quad 
    \p r \dot{\mu} = 0 , 
\label{d:r:omega:mu}
} 
where the last one holds for the radial null geodesics, 
for which $\mu = \pm 1$.

Now, suppose the universe is locally in thermal equilibrium,  
that is, $F$ is given as the Planck distribution function, $\Phi$, 
at the last scattering surface, as required in most of 
the known LTB models.   
Then, given a photon geodesic $\gamma$, 
the ratio of the temperature $T$ of the photon and its energy 
$\omega$ is preserved along the trajectory $\gamma$. 
Therefore, $\omega$ comes in $F$ in the form 
\Eq{
  F=\Phi(\omega/T) . 
} 
Then, the CMB temperature anisotropy $\delta T/T$ is generally given by 
\Eq{  
(\delta F)^{(1)} + (\delta F)^{(2)} + \cdots 
 = \left\{-\frac{\delta T}{T}\omega\p\omega 
 + \frac{1}{2}\left(\frac{\delta T}{T}\right)^2 (\omega\p\omega)^2 
 + \cdots \right\}\Phi . 
\label{eq: deltaF-deltaT} 
}
Now, suppose that an observer lives at a distance of $\delta x^i$ 
from the symmetry center. Then, the left-side are written as 
\Eq{
  (\delta F)^{(1)} = \delta x^i (\p i F)_0,
\ \ \ \ (\delta F)^{(2)} = \frac{1}{2}\delta x^i \delta x^j (\p i \p j F)_0,
}
where here and in the following, the subscript `$0$' implies 
the value evaluated at the center ($r = 0$) at the present time ($t = t_0$).  
The CMB temperature anisotropy dipole $(\delta T/T)^{(1)}$ 
and quadrupole $(\delta T/T)^{(2)}$ are therefore given by 
\Eqr{
 \left(\frac{\delta T}{T}\right)^{(1)} 
   &=& -\frac{\delta x^i (\p i F)_0}{\Fy}, \label{eq: dipole}\\
\left(\frac{\delta T}{T}\right)^{(2)} 
   &=& -\frac{1}{2}\frac{\delta x^i \delta x^j (\p i\p j F)_0}{\Fy}
       +\frac{1}{2}\left\{
                          \left(\frac{\delta T}{T}\right)^{(1)} 
                   \right\}^2 \frac{\Fyy}{\Fy}. 
\label{eq: quadrupole} 
}
Since (\ref{eq:F:spherical}) implies 
$\p i F = (\p i r)\Fr + (\p i \omega)\Fo + (\p i \mu)\Fm$,  
$(\p i F)_0$, $(\p i \p j F)_0$ in the right-hand side of 
the above equations are given by $\p\alpha F_0$ and $\p\alpha\p\beta F_0$ 
($\alpha, \beta = r, \omega, \mu$). 
Therefore, our task is to find the concrete expressions of 
$\p\alpha F_0$ and $\p\alpha\p\beta F_0$ in terms of 
relevant geometric quantities.

\subsection{The CMB dipole formula}
\label{subsec: dipole}

First, we will obtain the CMB temperature dipole formula. 
For this purpose, we derive the expression for $\p\alpha F_0$. 
Our stating point is the Boltzmann equation for $F_0(t,r,\omega,\mu)$,  
\Eq{
   \frac{d}{dt}F_0 = \Ft + \dot r \Fr + \dot{\omega}\Fo + \dot{\mu}\Fm 
                    = 0, 
\label{eq: Boltzmann} 
}
where ${\dot r}, {\dot \omega}, {\dot \mu}$ are defined for 
a given null geodesic curve $\gamma$. 
By differentiating this equation by $\alpha=r,\omega, \mu$, and 
using the formula (\ref{d:r:omega:mu}),  
we obtain the first order differential equations for $\partial_\alpha F$, 
\Eqr{
     \frac{d}{dt} \left( \begin{array}{c}
                 		\Fo \\
				\Fr \\
				\Fm \\
			 \end{array}
                  \right)
	= - \left( \begin{array}{ccc}
       		    \p\omega\dot{\omega} & 0 & 0 \\
		    \p r \dot{\omega} & \p r \dot r & 0 \\
		    \p\mu\dot{\omega} & \p\mu\dot r & \p\mu\dot{\mu} \\
                   \end{array}
			\right)
			\left( \begin{array}{c}
				\Fo \\
				\Fr \\
				\Fm \\
				\end{array}
			\right).
\label{eq: matrix}
}
The set of these equations can easily be integrated along 
the given photon trajectory $\gamma$ to yield the solutions 
\Eqr{
 \Fy &=& (\Fy)_i, \label{eq: Fy} \\
 \Fr &=& e^{-P(t, t_i)} (\Fr)_i 
         + (\Fy)_i \int^t_{t_i}dt_1 e^{-P(t, t_1)}A(t_1), \label{eq: Fr}\\
 \Fm &=& e^{-Q(t, t_i)} (\Fm)_i + (\Fr)_i \int^t_{t_i}dt_1 e^{-Q(t, t_1) 
         - P(t_1, t_i)}\left(-\frac{N}{S}\right)_{t_1} \nonumber\\
      && +(\Fy)_i \int^t_{t_i}dt_1 e^{-Q(t, t_1)}
                       \left\{
                              B(t_1) 
                             - \left(\frac{N}{S}\right)_{t_1}
                               \int^{t_1}_{t_i}dt_2 e^{-P(t_1, t_2)}A(t_2)
                       \right\}, \nonumber\\
	&& \label{eq: Fm}
}
where
\Eq{
 A \equiv \left( 
                \frac{\dot N}{N} \pm 2\frac{N'}{S} + \frac{\dot S}{S}
          \right)' , \ \ \ 
 B \equiv 2\left\{
             \frac{N'}{S} \pm \left(\frac{\dot R}{R}-\frac{\dot S}{S}\right)
           \right\},
\label{eq: AB}
}
and
\Eqr{	
 P(t_1, t_2) &\equiv & \mp\int^{t_2}_{t_1}dt\left(\frac{N}{S}\right)' , 
  \label{eq: P}\\
 Q(t_1, t_2) &\equiv & 2\int^{t_2}_{t_1}dt
                       \left\{\pm\frac{N}{S}
                                 \left(\frac{R'}{R} - \frac{N'}{N}\right) 
                               + \frac{\dot R}{R} - \frac{\dot S}{S} 
                      \right\} \nonumber\\
             &=& 2\left[\ln\frac{R}{N}\right]^{t_2}_{t_1} 
               + 2\int^{t_2}_{t_1}dt 
                  \left(\frac{\dot N}{N} - \frac{\dot S}{S} \right), 
\label{eq: Q}
}
and where the subscript `$i$' denotes the value evaluated 
at the last scattering surface.

Note that 
$\Fy$ is constant for any null geodesics, as in (\ref{eq: Fy}).  
Furthermore, as shown in Appendix, by inspecting the regularity of $F$ 
at the center, as well as the behavior of some relevant geometric quantities 
near the center, we can observe that $\partial_r F_0$ does not 
contribute to the leading behavior of $\partial_i F$ in the limit 
$r \rightarrow 0$. 
Therefore, we only need to find the expression of $\partial_\mu F$.

Using the null geodesic equation for a radial geodesic, $\mu = \pm 1$, 
$dt/dr = \pm S/N$, 
we find that \refeq{P} and \refeq{Q} become 
\Eq{
  P(t_1, t_2) = \left[\ln\frac{S}{N}\right]_{t_1}^{t_2} + U(t_1, t_2),
	\ \ \ \ 
  Q(t_1, t_2) = 2\left[\ln\frac{R}{N}\right]_{t_1}^{t_2} + 2U(t_1, t_2),  
\label{eq: PQ2}
}
where 
\Eq{
 U(t_1, t_2) \equiv \int^{t_2}_{t_1}dt 
                    \left(\frac{\dot N}{N} - \frac{\dot S}{S}\right). 
\label{eq: U}
}
Substituting these into \refeq{Fm}, we have 
\Eqr{
 \Fm &=& \frac{R^2}{N^2}\frac{N_i^2}{R_i^2}\e2U{t}{t_i}(\Fm)_i
      - \frac{R^2}{N^2}\frac{N_i}{S_i}(\Fr)_i 
        \int^t_{t_i}dt_1 \left(\frac{NS}{R^2}\right)_{t_1}e^{-2U(t, t_1) 
      - U(t_1, t_i)} \nonumber\\
     & & +\frac{R^2}{N^2}(\Fy)_i 
                \int^t_{t_i}dt_1 
                \left(\frac{N^2}{R^2}\right)_{t_1}\e2U{t}{t_1}\left\{B(t_1) 
              - \int^{t_1}_{t_i}dt_2 \eU{t_1}{t_2}A(t_2)\right\}. \nonumber\\
      && 
\label{eq: Fm2}
}
Here we note that the second and third terms in the right-side of the above 
equation have the form 
\Eq{
 I(r) \equiv \int^t_{t_i}dt_1 \frac{V(t, t_1)}{R^2(t_1)} 
 = \pm\int^r_{r_i}dr_1 
                  \left(\frac{S}{N}\right)_{r_1}\frac{V(r, r_1)}{R^2(r_1)},
}
where $V(r, r_1)$ is a function of $r$ and $r_1$ that is regular 
at $0 \le r \le r_1$. For the radial geodesic that reaches  
$r = 0$ at $t = t_0$, $I(r)$ behaves as 
\Eq{
  I(r) \simeq -\frac{1}{r}\frac{S_0}{N_0}\frac{V(0,0)}{(R'_0)^2}.
}
Substituting the above expression of $\Fm$ into 
\Eq{
 \p i F \simeq S_0 \frac{p^i}{p}(f_{\nu})_0 
 = S_0 \frac{p^i}{p}\left(\frac{\Fm}{r}\right)_0 ,\quad \mbox{as $r \to 0$} , 
}
which is derived in Appendix (see \refeq{Fi2}), we have 
\Eq{
 (\p i F)_0 \simeq \mp\frac{S_0^2}{N_0}\frac{p^i}{p} 
          \left[\frac{N_i}{S_i}\eU{t_0}{t_i}(\Fr)_i 
                + (\Fy)_i \left\{\int^{r_i}_0 dr \eU{t_0}{t}A(t)-B_0 \right\}
          \right].  
}
Thus, using this expression, we can write \refeq{dipole} as 
\Eq{
   \left(\frac{\delta T}{T}\right)^{(1)} 
   = \mp\frac{\delta L\mbox{\boldmath $n$}\cdot\Omega}{N_0}
     \left[
            \frac{N_i}{S_i}\eU{t_0}{t_i}\left(\frac{\Fr}{\Fy}\right)_i 
          + \int^{r_i}_0 dr \eU{t_0}{t}A(t) - B_0 
     \right],
\label{eq: dipole2}
}
where $\delta L\mbox{\boldmath $n$}$ is the position vector of the observer. 
This is our dipole formula for the CMB temperature anisotropy 
in the most general spherically symmetric spacetime. 
The regularity of the metric, (\ref{eq: spherically sym.}) at 
the symmetry center implies that 
$N^2 \simeq C_1 + \Ord{r^2}$, $R/r \simeq C_2 + \Ord{r^2}$, 
and $S^2 \simeq (R/r)^2 + \Ord{r^2}$ near the center, with 
$C_1,\:C_2$ being some constants with respect to $r$. 
Using these estimations, we can check that the right-hand side 
of \refeq{dipole2} is convergent, hence well-defined.

\subsection{The CMB quadrupole formula}
Next, we will derive the CMB quadrupole formula by inspecting 
the second-order derivatives, 
$\partial_\alpha \partial_\beta F_0$, $(\alpha, \beta = r, \omega, \mu)$.  
By differentiating the first row of \refeq{matrix} with respect to 
$\ln\omega$, $r$, and $\mu$, we obtain 
\Eqr{
	\frac{d}{dt}\{\Fyy\} &=& 0, \\
	\frac{d}{dt}(\Fyr) &=& \mp\left(\frac{N}{S}\right)' \Fyr + A\Fyy, \\
	\frac{d}{dt}(\Fym) &=& 2\left\{\pm\frac{N}{S}\left(\frac{R'}{R} -\frac{N'}{N}\right) + \frac{\dot R}{R} - \frac{\dot S}{S}\right\}\Fym \nonumber\\
								&& -\frac{N}{S}\Fyr + B\Fyy,
}
for a radial geodesic for which $\mu = \pm 1$.
We can easily integrate this set of equations, and get the solutions 
\Eqr{
 \Fyy &=& \{\Fyy\}_i , \\
 \Fyr &=& e^{-P(t, t_i)}(\Fyr)_i 
         + \{\Fyy\}_i\int^t_{t_i}dt_1 e^{-P(t, t_1)}A(t_1), \label{eq: Fyr}\\
 \Fym &=& e^{-Q(t, t_i)}(\Fym)_i 
          - \{\Fyr\}_i\int^t_{t_i} \! dt_1 e^{-Q(t, t_1) 
          - P(t_1, t_i)}\left(\frac{N}{S}\right)_{t_1} \nonumber\\
      && +\{\Fyy\}_i \int^t_{t_i} \! dt_1 e^{-Q(t, t_1)}  
                      \left\{B(t_1) 
                            - \left(\frac{N}{S}\right)_{t_1}
                              \int^{t_1}_{t_i}  \! dt_2 e^{-P(t_1, t_2)}A(t_2)
                      \right\}. \nonumber\\
      &&
}
Similarly, we can get the ordinary differential equations
\Eqr{
 \frac{d}{dt}(\Frr) 
     &=& \mp 2\left(\frac{N}{S}\right)' \Frr 
         \mp \left(\frac{N}{S}\right)'' \Fr + A' (\Fy)_i + 2A\Fyr, \\
 \frac{d}{dt}(\Frm) 
     &=& \left[\mp\left(\frac{N}{S}\right)'
               + 2\left\{\pm\frac{N}{S}\left(\frac{R'}{R}-\frac{N'}{N}\right) 
                         + \frac{\dot R}{R} - \frac{\dot S}{S}
                  \right\}
         \right]\Frm 
       - \frac{N}{S}\Frr \nonumber\\
     && +B\Fyr + A\Fym - \left(\frac{N}{S}\right)' \Fr + C\Fm 
        + B' (\Fy)_i , \nonumber\\
     && \\
 \frac{d}{dt}(\Fmm) 
     &=& 4\left\{\pm\frac{N}{S}\left(\frac{R'}{R} -\frac{N'}{N}\right) 
         + \frac{\dot R}{R} - \frac{\dot S}{S}\right\}\Fmm 
         - 2\frac{N}{S}\Frm +2B\Fym \nonumber \\
     && +D\Fm + 2\left(\frac{\dot S}{S} - \frac{\dot R}{R}\right)(\Fy)_i,
}
for a radial geodesic, $\mu =\pm 1$.
We can also integrate this set of equations, and obtain the solutions 
\Eqr{
 \Frr &=& e^{-2P(t, t_i)}(\Frr)_i \nonumber\\
       && + \int^t_{t_i}dt_1 e^{-2P(t, t_1)}
                 \left\{\mp\left(\frac{N}{S}\right)'' \Fr 
                        + A' (\Fy)_i + 2A\Fyr 
                 \right\}_{t_1}, \label{eq: Frr}\\
 \Frm &=& e^{-P(t, t_i) - Q(t, t_i)}(\Frm)_i \nonumber\\
       && + \int^t_{t_i}dt_1 e^{-P(t, t_1) 
                   - Q(t, t_1)}
            \Biggl\{-\frac{N}{S}\Frr +B\Fyr  + A\Fym \nonumber\\
       &&\hspace{121pt} - \left(\frac{N}{S}\right)'\Fr  + C\Fm + B' (\Fy)_i  \Biggr\}_{t_1}, \\
 \Fmm &=& e^{-2Q(t, t_i)}(\Fmm)_i 
         + \int^t_{t_i}dt_1 e^{-2Q(t, t_1)}
           \Biggl\{-2\frac{N}{S}\Frm +2\left(\frac{\dot S}{S} - \frac{\dot R}{R}\right) \! (\Fy)_i \nonumber\\
       && \hspace{172pt} + 2B\Fym + D\Fm\Biggr\}_{t_1}, 
} 
where
\Eq{
    C \equiv 2\left\{\pm\frac{N}{S}\left(\frac{R'}{R} -\frac{N'}{N}\right) 
         + \frac{\dot R}{R} - \frac{\dot S}{S}\right\}' , \ \ \ 
    D \equiv 2\frac{N}{S}\left(\frac{R'}{R} -\frac{N'}{N}\right) 
             \pm 6\left(\frac{\dot R}{R} - \frac{\dot S}{S}\right).
}

Thus, we have the six solutions, $(\omega\partial_\omega)^2F_0$, 
$\omega \partial_\omega \partial_r F_0$, etc. However, 
we note that $(\omega\partial_\omega)^2F_0$ is just constant. 
Furthermore, by inspecting the regularity of $F_0$ at the center, 
as well as the behavior of some geometric quantities near the center, 
we can find that only $\partial_r^2 F_0$ becomes relevant to the evaluation 
of $\partial_i \partial_j F$ in the limit $r \rightarrow 0$. 
In fact, as we show in Appendix~(see (\ref{eq: Fij_app})),
\Eqr{
   \p i \p j F &\to &
            2 \left(
                    \delta_{ij} - S_0^2 \frac{p^i p^j}{p^2} 
              \right)(f_2)_0 
           + S_0^2 \frac{p^i p^j}{p^2}(\Frr)_0 \nonumber\\
         &&  + \left\{
                    \left(
                          \frac{\ape''}{\ape} - \frac{N''}{N} 
                    \right)_0 \delta_{ij} 
                 + C_0 \frac{p^i p^j}{p^2}  
             \right\}(\Fy)_i , 
\label{eq: Fij}
}
where $f_2 = \partial_{r^2}F$ (see (\ref{eq: f}) in Appendix), and  
$a_\perp \equiv R/r$, which corresponds to the `scale factor' 
perpendicular to the radial direction. 
Then, in terms of $f_2$ and $\partial_r^2 F_0$, 
the CMB quadrupole formula (\ref{eq: quadrupole}) is written as 
\Eqr{
     \left(\frac{\delta T}{T}\right)^{(2)} 
     &=& -\frac{1}{2}\frac{\delta x^i \delta x^j}{(\Fy)_i}
	  \Biggl[
                 2(\delta_{ij} - \Omega_i \Omega_j )(f_2)_0 
               + \Omega_i \Omega_j (\Frr)_0 \nonumber\\
      &&       \hspace{60pt} +\left\{\frac{\ape''}{\ape}\delta_{ij}
                 +\ape\left(S'' - \ape''\right)
                  \frac{\Omega_i \Omega_j}{S^2}\right\}_0 (\Fy)_i 
          \Biggr] \nonumber\\
      && +\frac{1}{2}\left\{ 
                            \left(\frac{\delta T}{T}\right)^{(1)}
                     \right\}^2 \frac{\{\Fyy\}_i}{(\Fy)_i}, 
\label{eq: quadrupole2}
}
where $\Omega_i \equiv \delta_{ij}x^j/r$. 
So, the remaining task is to find the expressions of the leading 
behavior of $f_2$ and $\Frr$ at the center $r \to 0$.

First, we note from \refeq{f2} that $(f_2)_0$ is given by 
\Eq{
(f_2)_0 = \frac{1}{2}\left(\frac{\Fr}{r} \mp \frac{\Fm}{r^2}\right)_0 .
\label{eq: f2-2}
}
In the limit $r \to 0$, from \refeq{Fm}, the second term of this equation 
can be written as 
\Eqr{
\frac{\Fm}{r^2} 
   &\to & \frac{{\ape}_0^2}{N_0^2}\frac{N_i^2}{R_i^2}\e2U{t_0}{t_i}(\Fm)_i 
             \mp\frac{{\ape}_0^2}{N_0^2}
             \int^{r_i}_0 \!\! dr \frac{N^2}{r^2}\frac{\e2U{t_0}{t}}{\ape^2}
                             \left( \! -\Fr + \frac{S}{N}B\Fy\right) \nonumber\\
   &=& \frac{{\ape}_0^2}{N_0^2}\frac{N_i^2}{R_i^2}\e2U{t_0}{t_i}(\Fm)_i
       \pm\frac{{\ape}_0^2}{{\ape}^2_i}\frac{\e2U{t_0}{t_i}}{r_i}
                             \left(-\Fr + \frac{S}{N}B\Fy\right)_i \nonumber\\
   && \mp\left\{\frac{1}{r}\left(-\Fr + \frac{S}{N}B\Fy\right)\right\}_0 
\nonumber\\
   && \pm\frac{{\ape}_0^2}{N_0^2}\int^{t_0}_{t_i}dt 
         \frac{N^2}{r}\frac{d}{dt}
         \left\{\frac{\e2U{t_0}{t}}{\ape^2}
                \left(-\Fr + \frac{S}{N}B\Fy\right)
         \right\}.
\label{eq: Fm/r^2}
}
From the second row of \refeq{matrix}, we have
\Eqr{
\frac{d}{dt}\left(\frac{\Fr}{\ape^2}\e2U{t_0}{t}\right)
 &=& \left\{
             2\left(
                     \frac{\dot S}{S} - \frac{\dot R}{R}
                   - \frac{\dot N}{N} \mp \frac{N}{S}\frac{\ape'}{\ape}
              \right) 
              \mp\left(\frac{N}{S}\right)'
            \right\}\frac{\Fr}{\ape^2}\e2U{t_0}{t} \nonumber\\
 && +\frac{A}{\ape^2}\e2U{t_0}{t}(\Fy)_i 
      .
}
Thus, $(f_2)_0$ becomes
\Eqr{
(f_2)_0 &=& \mp\frac{1}{2}\frac{{\ape}_0^2}{N_0^2}
                         \frac{N_i^2}{R_i^2}\e2U{t_0}{t_i}(\Fm)_i
            + \left(\frac{1}{r}\frac{S}{N}B\Fy\right)_0 \nonumber\\
          && 	    - \frac{{\ape}_0^2}{{\ape}^2_i}
              \frac{\e2U{t_0}{t_i}}{2r_i}
              \left(-\Fr + \frac{S}{N}B\Fy\right)_i \nonumber\\
   && -\frac{1}{2}\frac{{\ape}_0^2}{N_0^2}\int^{t_0}_{t_i}\!\! dt \frac{N^2}{r}
      \Biggl[
             -\left\{
                    2\left(
                            \frac{\dot S}{S} - \frac{\dot R}{R}
                            - \frac{\dot N}{N} 
                          \mp \frac{N}{S}\frac{\ape'}{\ape}
                     \right) + \left(\frac{N}{S}\right)'
              \right\}\Fr \nonumber\\
   && \hspace{88pt} +\left\{-A 
                           - 2\left(
                                    \frac{\dot N}{N} - \frac{\dot S}{S}
                              \right)\frac{S}{N}B \right. \nonumber\\
	&& \hspace{108pt} \left. + \ape^2 \frac{d}{dt}\left(\frac{SB}{N\ape^2}\right)
                     \right\}(\Fy)_i 
       \Biggr]\frac{\e2U{t_0}{t}}{\ape^2}. 
\label{eq: f2-3}
}
Next, from \refeq{Frr}, we obtain
\Eqr{
 (\Frr)_0 &=& e^{-2P(t_0, t_i)}(\Frr)_i \nonumber\\
				&& + \int^{t_0}_{t_i}dt e^{-2P(t_0, t)}\left\{2A\Fyr + A' (\Fy)_i \pm \left(\frac{N}{S}\right)'' \Fr\right\}. \ \ \ 
\label{eq: Frr2}
}

Thus, substituting \refeq{f2-3} and \refeq{Frr2} with \refeq{Fr} and \refeq{Fyr} into \refeq{quadrupole2}, 
we finally obtain the quadrupole formula for the CMB temperature anisotropy.
As in the dipole formula case, under the assumption that our metric 
\refeq{spherically sym.} is regular at the symmetry center, we can 
check that the above quadrupole formula is well-defined.

\section{The CMB temperature anisotropy in the LTB model} 
\label{sec: LTB}

In this section, the CMB temperature anisotropy formulae 
obtained in the previous section will be given concrete expressions 
for the LTB cosmological models, and then will be applied 
in some specific LTB models considered in the numerical analyses of 
Alnes and Amarzguioui \cite{AA}, to check the consistency of the formulae.  
Further, the constraint on the location of off-center observers 
will be derived by our analytic formulae with the latest WMAP data, 
thereby updating the previous results numerically obtained. 
But before doing so, we will first briefly recapitulate the LTB metric.

\subsection{The LTB spacetime}
\label{subsec: LTB}

A spherically symmetric spacetime with only non-relativistic matter, 
or dust, is described by the LTB metric, which may be given by 
setting in (\ref{eq: spherically sym.}), 
\Eqr{
N^2 = 1,\quad S= \frac{R'(t,r)}{1 - k(r)r^2} ,
}
with $k(r)$ begin an arbitrary function of $r$.
Then, the Einstein equations reduce to 
\Eq{
	\left(\frac{\dot R}{R}\right)^2 = \frac{2GM(r)}{R^3} - \frac{k(r)r^2}{R^2}, \label{eq: Einstein1}
	\ \ \ \ 4\pi\rho (t,r) = \frac{M' (r)}{R^2 R'},
}
where $M(r)$ is an arbitrary function of only $r$, and $\rho(t, r)$ is 
the energy density of the dust fluid. 
The solutions to \refeq{Einstein1} depend on the sign of $k(r)$ 
and can be expressed in parametric form: 
For $k(r)>0$, we have
\Eq{
  R(t, r) = \frac{M(r)}{k(r)r^2}(1 - \cos\eta),	\ \ \ \ 
  t - t_s (r) = \frac{M(r)}{\{k(r)r^2 \}^{\frac{3}{2}}}(\eta - \sin\eta),
}
where $t_s(r)$ is an arbitrary function of only $r$. 
For $k(r)=0$, we have
\Eq{
  R(t, r) = \left(\frac{9}{2}\right)^{\frac{1}{3}}
            M^{\frac{1}{3}}(r)\{t - t_{s}(r)\}^{\frac{2}{3}}. 
}
For $k(r)<0$, we have 
\Eq{
    R(t, r) = \frac{M(r)}{-k(r)r^2}(\cosh\eta - 1), \ \ \ \ 
    t - t_s (r) = \frac{M(r)}{\{-k(r)r^2 \}^{\frac{3}{2}}}(\sinh\eta - \eta). 
}
The area radius $R(t, r)$ vanishes at $t = t_s(r)$, 
so that $t_s(r)$ is called the big-bang time. 
The solutions admit three arbitrary functions $k(r)$, $M(r)$ and $t_s(r)$, 
but due to one degree of freedom in rescaling $r$, only two of them 
are independent. By appropriately choosing the profile of these two 
arbitrary functions, one can construct LTB cosmological models 
that can reproduce the observed SNIa distance-redshift relation.

\subsection{The analytic formula for the CMB dipole}
\label{subsec: dipoleLTB}
In the LTB cosmological model, in addition to $a_\perp = R/r$ defined 
previously, we also 
introduce the `scale factor along the radial direction' 
by $\apa(t, r) \equiv R'(t, r)$.  
Accordingly, we also define two Hubble expansion rates 
in the radial and azimuthal direction, respectively, by 
\Eqr{
 \Hpa \equiv \dfrac{\dot{S}}{S} 
           = \frac{\dot{a}_{{\scriptscriptstyle /\! /}}}{\apa} , \quad 
 \Hpe \equiv \dfrac{\dot{R}}{R} 
           = \frac{\dot{a}_{\perp}}{\ape} .  
 } 

From \refeq{AB} and \refeq{U}, we obtain 
\Eq{
	U(t_1, t_2) = -\int^{t_2}_{t_1}dt \Hpa,
	\ \ \ \ A = \Hpa' ,
	\ \ \ \ B = \pm 2(\Hpa - \Hpe).
}
Then, the analytic formula for the CMB temperature 
anisotropy dipole \refeq{dipole2} takes the form 
\Eq{
    \left(\frac{\delta T}{T}\right)^{(1)} 
  = \mp\delta L\mbox{\boldmath $n$}\cdot\Omega
       \left\{
              \frac{\eU{t_0}{t_i}}{S_i}
              \left(\frac{\Fr}{\Fy}\right)_i 
            + \int^{r_i}_0 dr \Hpa' \eU{t_0}{t} 
       \right\}.
\label{eq: dipoleLTB}
}

Now, using this formula, we derive some constraints concerning 
the position of off-center observers. 
In general, the CMB temperature anisotropy can be decomposed in terms 
of the spherical harmonics $Y_{lm}$ by 
$\delta T/T = \sum_{l, m} a_{lm}Y_{lm}$. 
We are interested in $a_{10}$ as the dipole moment.
From \refeq{dipoleLTB}, we obtain
\Eq{
 a_{10} = \mp\sqrt{\frac{4\pi}{3}}\delta L
          \left\{
                 \frac{\eU{t_0}{t_i}}{S_i}
                 \left(\frac{\Fr}{\Fy}\right)_i 
               + \int^{r_i}_0 dr \Hpa' \eU{t_0}{t}
          \right\}.
}
Assuming that the universe is locally in thermal equilibrium at the last scattering surface, the first term in the bracket is of the order of $(\p r T/T)_i$ because $F_0$ is isotropic and depends only on $\omega/T_i$. In the models we consider in the present paper, this term can be neglected because the void size is sufficiently smaller than the horizon size and therefore, the observed region of the LTB universe is homogeneous with good accuracy on the last scattering surface. Furthermore, in order to estimate this term correctly, we have to specify the universe model before the last scattering. This is beyond the scope of this paper. Therefore, we have only estimated the contribution of the second term numerically. For the LTB model considered in \citen{AA}, we have found that the induced $a_{10}$ is about $1.23\times 10^{-3}$ or less, according to WMAP data \cite{alm}, which implies that the distance from the observer to the center, $\delta L$, has to be, $\delta L \lineq 16 \rm Mpc$.  This is consistent with the numerical result of \citen{AA}. We have also applied this formula to various LTB models, and found, for example, $\delta L \lineq 14 \rm Mpc$ in the Garfinkle model~\cite{Garfinkle}, and $\delta L \lineq 12 \rm Mpc$ in the GBH model~\cite{GBH}. 

\subsection{The analytic formula for the CMB quadrupole}
\label{subsec: quadrupoleLTB}
As for the quadrupole moment in the LTB model, 
from \refeq{f2-3} and \refeq{Frr2}, we obtain 
\Eqr{
   (f_2)_0 &=& \mp\frac{1}{2}\frac{a_0^2}{R_i^2}\e2U{t_0}{t_i}(\Fm)_i
                -\frac{a_0^2}{{\ape}^2_i}\frac{\e2U{t_0}{t_i}}{2r_i}
                 \left(-\Fr + \frac{\apa}{\xi} B\Fy\right)_i
\nonumber\\
	&& -\frac{a_0^2}{2}\int^{t_0}_{t_i}dt \frac{1}{r}
 	    \Biggl[
                   -\left\{
                          2\left(\Hpa - \Hpe 
                           \mp \frac{\xi}{\apa}
                               \frac{\ape'}{\ape}\right) 
                             + \left(\frac{\xi}{\apa}\right)'
                    \right\}\Fr 
\nonumber\\
        && \hspace{68pt} 
          + \left\{-\Hpa' + 2\Hpa\frac{\apa}{\xi}B \right.\nonumber\\ 
         && \hspace{86pt} \left. +\ape^2 \frac{d}{dt}\left(\frac{SB}{N\ape^2}\right)
            \right\}(\Fy)_i 
            \Biggr]\frac{\e2U{t_0}{t}}{\ape^2}, \label{eq: f2LTB}\\
   (\Frr)_0 &=& e^{-2P(t_0, t_i)}(\Frr)_i \nonumber\\
				&& + \int^{t_0}_{t_i}dt e^{-2P(t_0, t)}  
              \left\{
                     2 \Hpa' \Fyr + \Hpa'' (\Fy)_i 
                     + \left(\frac{\xi}{\apa}\right)''\Fr 
              \right\}, \nonumber\\
          && \label{eq: FrrLTB} 
}
where $a_0 \equiv S_0 = {\apa}_0 = {\ape}_0$, and 
$\xi \equiv\sqrt{1 - k(r)r^2}$, just for notational simplicity, 
and where 
\Eq{
  P(t_1, t_2) = \mp\int^{t_2}_{t_1}dt 
                   \left(\frac{\xi}{\apa}\right)' ,    
\label{P:12}
}
which is obtained from \refeq{P}. 
Thus, we now have the analytic formula for the quadrupole moment of 
the CMB temperature anisotropy in the LTB model: 
\refeq{quadrupole2} together with \refeq{f2LTB} and \refeq{FrrLTB}.

From \refeq{quadrupole2}, we derive 
\Eqr{
 a_{20} &=& -\sqrt{\frac{16\pi}{45}}\frac{(\delta L)^2}{2a_0^2}
           \left\{
                  - \frac{2(f_2)_0}{(\Fy)_i} + \frac{(\Frr)_0}{(\Fy)_i} 
                  + \frac{(S'' - \ape'')_0}{a_0}
           \right\} \nonumber\\
	 && + \frac{(a_{10})^2}{2\sqrt{5\pi}}\frac{(\Fyy)_i}{(\Fy)_i}. 
}
If the universe is locally in thermal equilibrium at the beginning, 
we can set  $(\p\alpha\Fm)_i$ to be zero.  Further, for the same reason as we explained for the dipole formula, we  neglect the term  $(\p\alpha\Fr)_i$ in the present paper. 
Under these assumption, we have estimated the quadrupole moment using this formula numerically for the model in \citen{AA}, and found that $a_{20} \simeq 8.61\times 10^{-7}$. This is consistent with the numerical result of \citen{AA}. For other models, for example, $a_{20} \simeq 5.51\times 10^{-6}$ in the Garfinkle model \cite{Garfinkle}, and $a_{20} \simeq -9.27\times 10^{-7}$ in the GBH model~\cite{GBH}.

\section{Summary and discussions} 
\label{sec: summary}

In this paper, we have derived the analytic formulae for the dipole \refeq{dipole2} and quadrupole \refeq{quadrupole2} moments of the CMB 
temperature anisotropy in general spherically symmetric 
spacetimes, including the LTB cosmological model as a special case. 
The formulae can be used to compare consequences of 
the LTB/local void models with observations of the CMB temperature 
anisotropy rigorously. The formulae also enable us to identify physical origins 
of the CMB temperature anisotropy in the LTB models. 
For example, in the CMB dipole formula \refeq{dipole2}, 
the first term comes from the initial (spherical) inhomogeneity 
at the last scattering surface, while the second term represents 
the integrated Sachs-Wolfe effect. Note that the first term also contains a contribution that reduces to the second-order ISW effect in the spatially homogeneous case.

We have checked the consistency of our formulae for both 
dipole and quadrupole, with the widely-used recent numerical results for 
the special LTB model by H. Alnes and M. Amarzguioui\cite{AA}. Furthermore, we applied our formulae 
to other LTB models, such as those in \citen{Garfinkle, GBH} 
and in particular, for the dipole moment, we found the constraints 
on the distance between the void center and an off-center observer, by using the latest WMAP data.

We can also utilize our analytic quadrupole formula to discuss the relevance of the LTB model to observed anomalies. For example, the observed magnitude of the quadrupole is known to be significantly lower than the $\Lambda$CDM model predicts. This is usually understood as a cosmic variance, i.e., to be produced by a special feature of our Universe, one particular realization of the statistical ensemble. Because the local void model is one of such realization with a very low probability in the standard  $\Lambda$CDM model, it is tempting to see whether the quadrupole anomaly of the CMB anisotropy can be explained by a local void model.  Unfortunately, however, the above analysis of the constraint on the observer offset by the dipole moment implies that the observed anomaly cannot be explained solely by the induced quadrupole moment in LTB models such as those in \citen{AA,Garfinkle,GBH}.  Nevertheless, this result is not conclusive. For example, we have implicitly assumed that the off-center observer stays at a fixed comoving position. If the observer has a peculiar velocity pointed toward the center of the void, however, the value of $\delta L$ could be chosen to be much larger than the case 
with no peculiar velocity. If it is the case, then the observed anomaly of the quadrupole could be explained within the LTB models of \citen{AA,Garfinkle,GBH}. 
Therefore, it would also be worth attempting to develop  other analytic formulae concerning CMB polarizations, lensing effects, etc. (cf. \citen{GK}) that can be used to distinguish the LTB and FLRW cosmologies.

\section*{Acknowledgements}
We would like to thank Eiichiro Komatsu for useful information and Hajime Goto for discussions and comments. We also would like to thank all participants of the workshop $\Lambda$-LTB Cosmology (LLTB2009) held at KEK from 20 to 23  October 2009 for enlightening and joyful discussions. 
This work was supported in part by the project Shinryoiki of 
the SOKENDAI Hayama Center for Advanced Studies and 
the MEXT Grant-in-Aid for Scientific Research on Innovative 
Areas (No. 21111006). 

\appendix

\section{The regularity and derivatives of $F_0$ near the center}

In this appendix, we discuss the regularity of the distribution function $F$ 
at the symmetry center, and find which of $\partial_\alpha F$ (resp. 
$\partial_\alpha \partial_\beta F$), $\alpha, \beta = r,\omega,\mu$, 
become relevant in the leading behavior of the first- (resp. second-) 
order derivatives of $F$ in the limit $r \to 0$.

Mathematically, the distribution function can become singular at some radius including at the center. Furthermore, some authors concluded that a $C^{2-}$-class singularity of the metric should be allowed at the center in order to construct a model with accelerated expansion exactly at the center\cite{Vanderveld}. However, such a model can be easily made smooth by appropriate smoothing and the original singular model can be recovered as a limit such that the smoothing length approaches zero. Hence, the final formulae for the dipole and quadrupole anisotropy of CMB given in the present paper can be applied also to models with singularity in the metric or the distribution function if the results are finite. Hence, in this appendix, we assume that the metric and the distribution function are regular and smooth everywhere in Cartesian-type space coordinates on which the rotational symmetry group acts as on the standard Cartesian coordinates of the Euclidean space.

In such a coordinate system $\bm{x}=(x^i)$, the spatial part of the metric \refeq{spherically sym.}, denoted here by $g_{ij}$, can be written as 
\Eq{
 g_{ij} = S^2 \Omega_i \Omega_j + \ape^2 (\delta_{ij} - \Omega_i \Omega_j),
}
where $\Omega^i \equiv x^i/r$ and $\ape \equiv R/r$.  For any smooth and rotationally invariant function $h$ in this coordinate system, when it is  Taylor expanded as
\Eq{
h=h_0 + h_i x^i + h_{ij} x^i x^j + \cdots,
}
each coefficient $h_{ij\cdots}$ must be a rotationally invariant constant tensor. Hence, these coefficient tensors vanish for odd ranks and can be written as the sum of products of the Kronecker delta $\delta_{ij}$ for even ranks. This implies that $h$ is a smooth function of $r^2=\delta_{ij} x^i x^j$. 

Similarly, a smooth distribution function in this coordinate system can be Taylor expanded as
\Eq{
F(t,\bm{x},\bm{p})=b_0(t,\bm{p}) + b_i (t,\bm{p})x^i + b_{ij}(t,\bm{p})x^i x^j + \cdots,
}
where $\bm{p}=(p^i)$. When $F$ is rotationally invariant, each term on the right-hand side is rotationally invariant separately. This implies that each $b_{i_1\cdots i_l}$ is an ${\rm SO}(3)$ tensor depending only on the non-trivial vector $p^i$ and therefore, can be written as the sum of products of the Kronecker delta $\delta_{ij}$ and the vector $\bm{p}=(p^i)$. This implies that $F$ can be written
\Eq{
F(t,\bm{x},\bm{p})=\t f(t,x_i x^i, |p|, x_i p^i),
}
where $x_ix^i=r^2$ and $|p|=(\delta_{ij}p^i p^j)^{1/2}$. Here, from
\Eq{
p^2 \equiv g_{ij}p^i p^j = C(x_i p^i)^2 + \ape^2 \delta_{ij}p^i p^j ,
}
where $C$ is a smooth function defined by
\Eq{
C(t, r) \equiv \frac{S^2 - \ape^2}{r^2},
}
it follows that $|p|$ is a smooth function of $t$, $\omega=p$, $x_i p^i$ and $r^2$. Therefore, the regularity of $F$ at the center is equivalent to the condition that the corresponding function $F_0(t,r,\omega,\mu)$ can be written in terms of a smooth function $f$ with the four arguments $t,r^2, y \equiv \ln\omega,\nu \equiv r\mu= S x_i p^i/p$
\Eq{
    F_0 (t, r, \omega, \mu) = f(t, r^2 , y, \nu).
\label{eq: f}
}
Then, 
\Eq{
 \p i F = \p i f = \{\p i (r^2)\}f_2 + (\p i y)f_{y} + (\p i \nu)f_{\nu},
\label{eq: Fi} 
}
where $f_2 \equiv \p{r^2}f, f_y \equiv \p y f$, and $f_{\nu} \equiv \p\nu f$. Here, because the spatial derivative of $p^2$ can be written 
\Eq{
	\p i (p^2) = \left\{\frac{C'}{r}(rp^r)^2 + \frac{(\ape^2)'}{r}\delta_{jk}p^j p^k \right\}x^i + 2C(rp^r)p^i ,
}
we have 
\Eqr{
     \p i (r^2) &=& 2x^i , \\
     \p i (\ln\omega) 
           &=& \left\{
                  -\frac{N'}{rN} + \frac{\ape'}{r\ape} 
                  + \left(\frac{C'}{2r} 
                  - \frac{C\ape'}{r\ape}\right)\left(\frac{\mu r}{S}\right)^2 
               \right\}x^i 
            + C\frac{\mu r}{S}\frac{p^i}{p}, \\ 
     \p i \nu &=& S\frac{p^i}{p} + rS' \frac{x^i}{r}\frac{p^r}{p} 
               - rSp^r \frac{\p i (p^2)}{2p^3}. 
}
In particular, in the limit $r \to 0$, we see 
\Eqr{
	C {}\qquad {}\qquad {} &\to & \{\ape(S'' - \ape'')\}_0 , \\
	\p i (p^2) , \, \p i (r^2),\, \p i (\ln\omega) &\to & \Ord r, \\
	\p i \nu {}\qquad {} \qquad {} &\to & S_0 \frac{p^i}{p}.
}
Therefore, from \refeq{Fi}, we find that the first derivative of $F_0$ 
behaves at the center as 
\Eq{
    \p i F \; \to \; S_0 \frac{p^i}{p}(f_{\nu})_0 
               = S_0 \frac{p^i}{p}\left(\frac{\Fm}{r}\right)_0 \,.
\label{eq: Fi2}
}

Next we study the second order derivatives of $F$ with respect to $x^i$, 
which are written as 
\Eqr{
 \p i \p j F 
       &=& \{\p i \p j (r^2)\}f_2 
           + \{\p j (r^2)\}
                \left[ 
                     \{\p i (r^2)\}f_{22} + (\p i y)f_{2y} + (\p i \nu)f_{2\nu}
                \right]  \nonumber\\ 
       && +(\p i \p j y)f_y 
          + (\p j y)\left[\{\p i (r^2)\}f_{y2} 
                          + (\p i y)f_{yy} + (\p i \nu)f_{y\nu}  
                    \right] \nonumber\\
       && +(\p i \p j \nu)f_{\nu} 
          + (\p i \nu)\left[\{\p i (r^2)\}f_{\nu 2} 
                            + (\p i y)f_{\nu y} + (\p i \nu)f_{\nu\nu} 
                      \right]. 
}
We find that 
\Eqr{
	\p i \p j (r^2) &=& 2\delta_{ij}, \\
	\p i \p j (p^2) &=& \left[\left(\frac{C'}{r}\right)' (rp^r)^2 + \left\{\frac{(\ape^2 )'}{r}\right\}'\delta_{kl}p^k p^l \right]x^i \frac{x^j}{r} 
								+2\frac{C'}{r}(rp^r)(p^i x^j + p^j x^i )\nonumber\\
							&& +\left\{\frac{C'}{r}(rp^r)^2 + \frac{(\ape^2)'}{r}\delta_{kl}p^k p^l \right\}\delta_{ij} + 2Cp^i p^j, \\
	\p i \p j (\ln\omega) &=& \frac{\p i \p j (p^2)}{2p^2} - \frac{\p i (p^2)\p j (p^2)}{2(p^2)^2} - \left(\frac{N'}{rN}\right)' \frac{x^i}{r}x^j - \frac{N'}{rN}\delta_{ij}, \\
	\p i \p j \nu &=& \frac{S'}{r}\left[\frac{x^i p^j + 2x^j p^i}{p} - \frac{rp^r}{2p^3}\{x^j \p i (p^2) + x^i \p j (p^2)\}\right]
							+ \left(\frac{S'}{r}\right)' \frac{rp^r}{p}\frac{x^i x^j}{r} \nonumber\\
						&& -\frac{S}{2p^3}\{p^i \p j (p^2) + p^j \p i (p^2)\}
								- S\frac{rp^r}{2p^3}\left\{\p i \p j (p^2) - \frac{3}{2p^2}\p i (p^2) \p j (p^2)\right\}. \nonumber\\
						&&
}
In particular, in the limit $r \to 0$,
\Eqr{
	\p i \p j (r^2) &\to & 2\delta_{ij}, \\
	\p i \p j (p^2) &\to & 2\left(\frac{\ape''}{\ape}\right)_0 p^2 \delta_{ij} + 2C_0 p^i p^j, \\
	\p i \p j (\ln\omega) &\to & \left(\frac{\ape''}{\ape} - \frac{N''}{N}\right)_0 \delta_{ij} + C_0 \frac{p^i p^j}{p^2}, \\
	\p i \p j (\nu) &\to & \Ord r.
}
From these we find that in the limit $r \to 0$, 
\Eq{
	\p i \p j F \to 2\delta_{ij}(f_2)_0 + \left\{\left(\frac{\ape''}{\ape} - \frac{N''}{N}\right)_0 \delta_{ij} + C_0 \frac{p^i p^j}{p^2}\right\}(f_y)_0
							+ S_0^2 \frac{p^i p^j}{p^2}(f_{\nu\nu})_0 .
}
Now, from \refeq{f}, we find that 
$\Fr = 2rf_2 + \mu f_{\nu}$, $\Fm = rf_{\nu}$, $\Fy = f_y$, 
and $\Frr = 2f_2 + 4r^2 f_{22} + 4r\mu f_{2\nu} + \mu^2 f_{\nu\nu}$, 
and hence in the limit $r \to 0$, 
\Eqr{
	f_{\nu} &\to & \left(\frac{\Fm}{r}\right)_0, \\
	f_2 &\to & \frac{1}{2}\left(\frac{\Fr}{r} - \mu\frac{\Fm}{r^2}\right)_0, \label{eq: f2}\\
	f_{\nu\nu} &\to & \frac{1}{\mu^2}\left(\Frr - \frac{\Fr}{r} + \frac{\mu}{r^2}\Fm\right)_0 \nonumber\\
						&=& \frac{1}{\mu^2}\left(\Frr - 2f_2 \right)_0 .
}
Thus, we finally obtain 
\Eqr{
	\p i \p j F &\to & 2\left(\delta_{ij} - S_0^2 \frac{p^i p^j}{p^2}\right)(f_2)_0 + S_0^2 \frac{p^i p^j}{p^2}(\Frr)_0 \nonumber\\
					&&		+\left\{\left(\frac{\ape''}{\ape} - \frac{N''}{N}\right)_0 \delta_{ij} + C_0 \frac{p^i p^j}{p^2}\right\}(\Fy)_i .
\label{eq: Fij_app}
}


\end{document}